# Patient Digital Twins for Chronic Care: Technical Hurdles, Lessons Learned, and the Road Ahead

**Micheal P. Papazoglou[1]**, *Scientific Academy for Service Technology e.V. (ServTech), Potsdam, 14467, Germany*

**Bernd J. Krämer[2] Jr.**, *Scientific Academy for Service Technology e.V. (ServTech), Potsdam, 14467, Germany*

**Mira Raheem[3]**, *Faculty of Computers & Artificial Intelligence, Cairo University, Cairo, 12613 Egypt*

**Amal Elgammal[4] IV**, *Scientific Academy for Service Technology e.V. (ServTech), Potsdam, 14467, Germany*

***Abstract*—** Chronic diseases constitute the principal burden of morbidity, mortality, and healthcare costs worldwide, yet current health systems remain fragmented and predominantly reactive. Patient Medical Digital Twins (PMDTs) offer a paradigm shift: holistic, continuously updated digital counterparts of patients that integrate clinical, genomic, lifestyle, and quality-of-life data. We report early implementations of PMDTs via ontology-driven modeling and federated analytics pilots. Insights from the QUALITOP oncology study and a distributed AI platform confirm both feasibility and challenges: aligning with HL7 FHIR and OMOP standards, embedding privacy governance, scaling federated queries, and designing intuitive clinician interfaces. We also highlight technical gains, such as automated reasoning over multimodal blueprints and predictive analytics for patient outcomes. By reflecting on these experiences, we outline actionable insights for software engineers and identify opportunities, such as DSLs and model-driven engineering, to advance PMDTs toward trustworthy, adaptive chronic care ecosystems.

**Actionable Insights**

- **Prioritize interoperability over isolation.** The use of Patient Medical Digital Twins (PMDTs) depends on their ability to interconnect across hospitals, wearable devices, and clinical registries. Adherence to interoperability frameworks – such as HL7 FHIR, OMOP, and established biomedical ontologies –

---

- **Embed governance within system architecture.** Data governance should be treated as first-class design principle. Consent management, data provenance, and access control policies must be codified as machine-actionable artifacts, enabling automated compliance with regulatory frameworks such as GDPR and HIPAA, and ensurung accountability and trust in federated analytics environments.
- **Design for usability and adaptability.** Sustainable adoption of PMDTs by clinicians and patients requires demonstrable clinical value, delivered through intuitive dashboards, explainable AI, and scenario-based "what-if" simulation tools. These interfaces should evolve alongside clinical workflows and emerging technologies to support adaptive, evidence-based descison-making.

Chronic diseases account for nearly three-quarters of global healthcare costs and mortality. Yet most health systems remain reactive: patients are treated after deterioration[1] occurs; that's data are fragmented across institutions and devices, and care pathways are rarely tailored to individual needs. For clinicians, this results in incomplete information and limited decision support; for patients, it often leads to inconsistent follow-up, delayed interventions, and diminished quality of life.

At the same time, the digital health landscape is evolving at an unprecedented pace. Wearables, biosensors, and electronic health records now generate unprecedented volumes of data, while artificial intelligence (AI) promises predictive models for risk stratification and treatment optimization. Despite this progress, most systems remain siloed, non-interoperable, and fragile in terms of scalability, trust, and privacy[2]. What is missing is not more data or algorithms, but a *unifying software paradigm* that can integrate heterogeneous data sources, ensure semantic interoperability, and enable continuous, patient-centered care.

Digital twins (DTs), virtual representations of physical entities, have already transformed domains such as aerospace and manufacturing by enabling real-time monitoring, predictive analysis, and proactive interventions[3]. In healthcare, however, existing DT initiatives remain limited in scope. Organ- or disease-specific models, such as digital pancreas systems for diabetes or cardiovascular simulations, show promising results but fail to address multimorbidity and the systematic ecosystem in which care is delivered[4,5].

*Patient Medical Digital Twins (PMDTs)* extend this concepts by providing holistic, continuously updated digital representations of individual patients, spanning clinical, genomic, behavioral, and psychosocial data. Beyond modeling individuals, PMDTs act as connectors across the health ecosystem: linking hospital databases, wearable data streams, patient-reported outcomes, regulatory frameworks, and population-level analytics. This integration enables both personalized treatment and system-wide learning, facilitating proasactive and preventive care[4,6].

Our work on PMDTs advances this vision through three complementary strands. First, an ontology-driven framework modeled in OWL and validated in the QUALITOP oncology pilot showed the feasibility of querying multimodal patient data[7]. Second, a federated analytics platform enabled privacy-preserving machine learning across hospital datasets without centralizing sensitive data. Third, our ongoing research on model-driven engineering (MDE) and domain-specific languages (DSLs) aims to make PMDTs more configurable, analyzable, and scalable..

This article synthesizes *lessons learned from these early implementations*, emphasizing both technical hurdles and practical insights. Key topics include alignment with interoperability standards such as HL7 FHIR and OMOP, embedding consent, provenance, and access policies into governance-aware architectures, scaling federated queries across distributed infrastructures, and designing interfaces that enhance usability for clinicians and patients. Drawing from these experiences, we provide actionable insights for software engineers seeking to design trustworthy, adaptive digital twins for healthcare.

The remainder of this article is organized as follows. Section II reviews the promise and pitfalls of current digital twin efforts in healthcare. Section III introduces the PMDT ecosystem and its conceptual building blocks. Section IV discusses key engineering challenges and opportunities, followed by *lessons learned from ontology-based* *modeling and federated pilots*. We outline a research agenda on DSLs and MDE, advancing PMDTs toward personalized, sustainable care..

## DIGITAL TWINS IN HEALTHCARE: PROMISE AND PITFALLS

Digital twins (DTs) in healthcare have attracted increasing attention, supported by a growing body of reviews and strategic roadmaps[3, 4, 6, 8, 9]. Collectively, this literature underscores both the transformative potential of DTs and the persistent barriers that prevent their translation into clinical practice.

**Promise.** The anticipated benefits of DTs span multiple levels of the healthcare continuum. At the patient level, DTs enable personalized medicine by simulating treatment responses, forecasting disease progression, and integrating multimodal data into coherent profiles[4, 5]. In diabetes care, for example, artificial pancreas models demonstrate how continuous physiological signals can guide adaptive insulin therapy[10], while cardiovascular and stroke twins illustrate how computational models can inform diagnosis and anticipate disease trajectories [11, 12]. At the system level, DTs are increasingly being investigated for operational optimization, resource allocation, and population-level risk stratification[4]. Recent meta-reviews confirm that DTs could serve as catalysts for more proactive, preventive, and participatory healthcare[6, 8].

**Pitfalls.** Despite these advances, most healthcare DT implementations remain at the proof-of-concept stage. Several challenges persist:

- **Narrow scope:** The majority of DTs are organ- or disease-specific and fail to capture multimorbidity, despite its prevalence in clinical practice[1].
- **Interoperability gaps:** Most DTs do not standardize data integration across EHRs, wearables, and registries. Misalignment with standards like HL7 FHIR and OMOP remains a key obstacle[2, 8].
- **Privacy and governance.** DTs rely on continuous streams of sensitive patient data, but implementations often lack robust mechanisms for consent management, provenance tracking, and compliance with data protection regulations such as the GDPR and HIPAA[6].
- **Clinical usability.** Many DT tools are designed around technical capabilities rather than clinical workflows, resulting in steep learning curves and limited end-user acceptance.

**Implications.** These pitfalls reveal a gap between the potential of DTs and their operational maturity. Importantly, these are not only general barriers but challenges encountered directly in our own work. In the QUALITOP oncology pilot, interoperability issues emerged during the integration of multimodal patient data across hospital systems. Similarly, the federated analytics platform exposed tensions between privacy-preserving governance and scalable machine learning performance. Clinician feedback consistently emphasized the importance of intuitive, workflow-aligned interfaces to ensure that DT systems function as clinical assets rather than operational burdens in care delivery.

These experiences reinforce that moving beyond isolated pilots requires digital twins that are holistic, extensible, and trustworthy. This recognition informed the engineering principles underlying *Patient Medical Digital Twins (PMDTs)*: an ecosystem-level infrastructure that integrates modular ontologies, federated architectures, and governance-aware mechanisms to support patient-centered, preventive, and adaptive chronic care.

## ENGINEERING THE PMDT ECOSYSTEM

To move beyond isolated organ- or disease-specific digital twins, we introduce the concept of Patient Medical Digital Twins (PMDTs): holistic, continuously updated digital representations of patients embedded within the broader healthcare ecosystem. Unlike conventional digital twin approaches that model single physiological systems, PMDTs integrate heterogeneous data sources including clinical records, genomic profiles, wearable sensor data, psychosocial indicators, and patient-reported outcomes into a coherent, semantically interoperable framework.

Figure 1 illustrates the PMDT ecosystem. At its core, PMDTs are structured around ontology-driven modular blueprints encompassing entities such as Medical Stakeholder, Patient, Disease, Treatment, Treatment Performance, Medical Safety, and Medical Pathways. Each blueprint captures distinct but interconnected dimensions of patient care and system dynamics. The Patient Blueprint exemplifies multimodality, incorporating diverse PatientData types such as quality-of-life, nutrition, lifestyle behaviors, and psychosocial and psychological indicators. These extensions are grounded in established standards and ontologies, including HL7 FHIR, OMOP, CTCAE, and the Disease Ontology DO, ensuring semantic rigor while maintaining extensibility for emerging data modalities.

Concrete progress towards realizing the PMDT vision has been achieved through three complementary strands:

1. **Ontology-Driven Modeling and Validation**: Modular blueprints were formally modeled in OWL to represent multimodal patient data and answer competency questions derived from clinical personas[13]. Validation in the QUALITOP oncology pilot demonstrated both the technical feasibility and clinical relevance of ontology-based PMDT representations[7].
2. **Federated Analytics and Privacy-Preserving AI**: We developed and evaluated a federated analytics platform that enables predictive insights across distributed hospital datasets without centralizing sensitive data[14]. This work demonstrated the practical viability of GDPR- and HIPAA-compliant analytics, highlighting how PMDTs can operationalize privacy-preserving computation within federated infrastructures.
3. **Model-driven Engineering and DSLs**: Building upon the PMDT foundation, we designed the Digital Twin Processing Language (DTPL), a domain-specific language embedded within a model-driven engineering (MDE) framework. DTPL enables clinicians, researchers, and data stewards to specify analytic workflows, consent rules, and treatment pathways at a high level of abstraction, which are subsequently automatically transformed into executable system components[15].

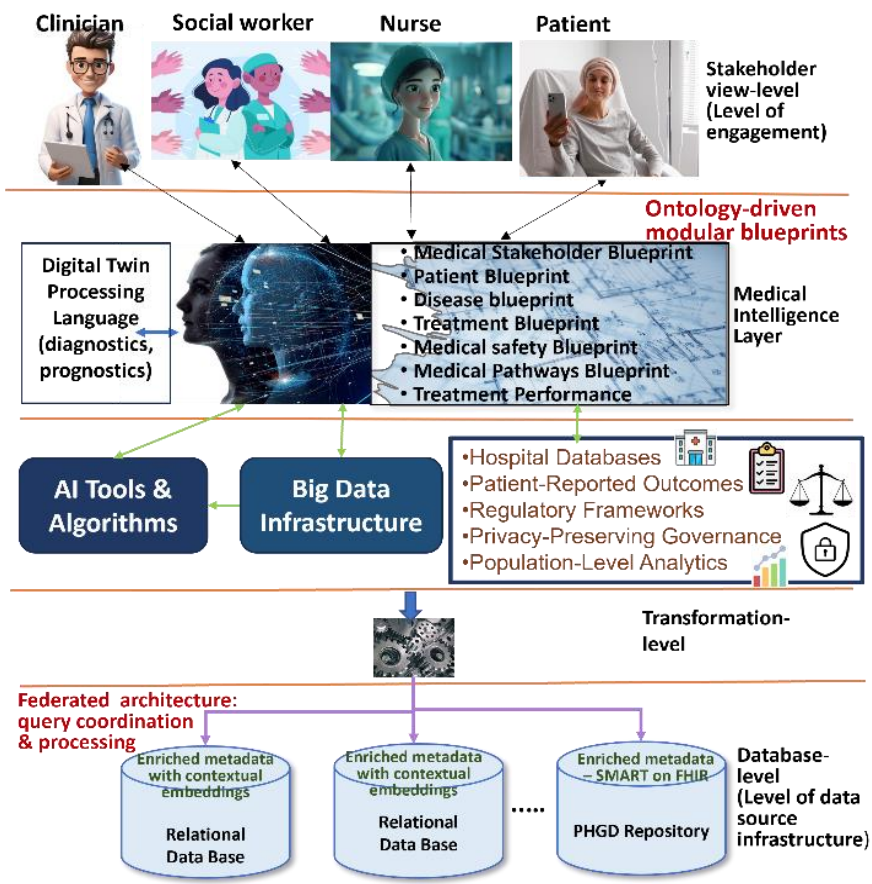

*Figure 1 The PMDT ecosystem vision. Ontology-driven modular blueprints integrate multimodal data sources and are processed through AI-enabled and big-data infrastructures. PMDTs interconnect with hospital databases, patient-reported outcomes, regulatory frameworks, and population-level analytics, enabling personalized treatment and system-wide learning under privacy-preserving governance.*

Collectively, these three strands show that PMDTs are not only a conceptual framework but an evolving ecosystem of interoperable methods and tools. This article emphasizes the ecosystem perspective and synthesizes lessons learned from early implementations, while companion studies provide detailed technical details of ontology-driven modeling and validation, federated analytics, and DSL components underpinning PMDT development.

Beyond individual care, *ecosystem integration* ensures that PMDTs act as connectors across hospital databases, clinical registries, wearable devices, and patient-reported outcomes. By embedding standards such as HL7 FHIR and OMOP, PMDTs integrate seamlessly with existing IT infrastructures rather than duplicating them. Privacy-preserving governance is embedded directly into the architecture through federated query mechanisms, explicit consent statements, and fine-grained access policies, ensuring that every data element remains traceable to its provenance and compliant with applicable regulatory frameworks.

Furthermore, PMDTs operate across multiple scales of granularity. At the *individual level*, they function as continuously updated companions, supporting shared decision-making, personalized treatment planning, and longitudinal monitoring. At the *ecosystem level*, aggregated PMDTs enable *learning health systems*, wherein federated analytics inform clinical guidelines, optimize resource allocation, and support population-level prevention strategies – all without compromising patient privacy and data sovereignty.

By uniting modular ontology-driven blueprints, ecosystem integration, federated analytics, and governance-aware architectures, PMDTs are evolving from a conceptual construct into a systematically engineered paradigm for trustworthy, adaptive, and patient-centered chronic care. This evolution positions PMDTs not only as technological artifacts but as foundational components of next-generation digital health ecosystems.

## ENGNEERING CHALLENGS AND OPPORTUNITIES

Realizing the vision of Patient Medical Digital Twins (PMDTs) requires more than technological optimism. It demands deliberate, system-level engineering to overcome barriers of interoperability, governance, scalability, usability, and maintainability. As illustrated in Figure 1, these challenges span all layers of the ecosystem -from federated data sources and orchestration mechanisms to the blueprint-based core and stakeholder-facing applications. Our experience shows that each challenge also represents an opportunity for the software engineering community to contribute methods, abstractions, and tools that ensure PMDTs remain sustainable and trustworthy at scale.

**Interoperability across Heterogeneous Standards:** Healthcare data remain fragmented across hospital systems, registries, wearable devices, and patient self-reporting platforms, each using different formats and standards such as HL7 FHIR, OMOP CDM, or proprietary APIs. In the QUALITOP pilot, ontology-driven alignment with standards proved essential for semantic consistency, while modular APIs enabled integration across distributed hospital datasets. For software engineers, this represents a classic systems integration problem: requiring semantic adapters and standardized interfaces that preserve scalability, fault tolerance, and data fidelity across heterogeneous infrastructures.

**Privacy, Provenance, and Governance:** PMDTs depend on the continuous processing of sensitive patient data. In our federated analytics implementation, GDPR compliance required embedding consent management, GDPR notices, provenance tracking, and access control policies directly within the architecture. This experience highlights the need for governance-aware models capable of automatically reasoning about which data may be shared, under what conditions, and with which stakeholder. Designing governance as a core architectural layer, rather than an external constraint, opens opportunities for developing trust-preserving infrastructures that integrate legal, ethical, and computational principals within a unified framework.

**Scalability and adaptability:** While a single PMDT may encapsulate a patient's multimodal health record, a fully realized PMDT ecosystem must scale across populations and remain adaptable to new therapies, biomarkers, and data modalities. Our distributed platform showed the feasibility of executing predictive queries within federated environments but also revealed open challenges related to versioning, schema evolution, and plug-and-play integration of new blueprints. Addressing these challenges demands advances in both ontology engineering and distributed software architectures, particularly around maintainability and evolution of large-scale knowledge-intensive systems.

**Human-Centered Design and Usability:** Even the most technically advanced PMDT will fail without adoption by clinicians and patients. Feedback from pilot evaluations confirmed that usability and workflow integration are critical barriers. Clinicians expect transparent, interpretable decision support without cognitive overload, while patients expect empowerment without technical complexity. These findings underscore the need for co-design approaches involving software engineers, human-computer interaction HCI specialists, and clinical stakeholders to ensure that PMDT interfaces are intuitive, explainable, and seamlessly integrated into everyday care workflow.

**Domain-Specific Languages and Model-Driven Engineering:** Managing the complexity of PMDT ecosystem requires abstractions that make them configurable, analyzable, and evolvable by design. Building upon our model-driven engineering (MDE) framework, we developed the *Digital Twin Processing Language (DTPL)* – a domain-specific language that allows high-level specification of patient blueprints, consent rules, and analytic workflows. In practice, DTPL supports three classes of operations: (i) *querying*, including both descriptive operations (e.g., for cross-correlating lifestyle factors across federated PMDTs) and *analytical* operations (e.g., for predicting adverse event probabilities); (ii) *composition and evolution*, combining blueprints to represent multimorbidity or integrating new patient-generated data streams; and (iii) *simulation and experimentation*, executing in silico tests and projecting treatment outcomes under alternative scenarios. By abstracting these operations, DTPL elevates PMDTs from static repositories to programmable, living entities that evolve systematically with clinical knowledge, therapeutic practice, and patient needs.

**Summary:** Our collective experience confirms that engineering PMDTs requires contributions from multiple subfields of software engineering: semantic interoperability, governance-aware architectures, distributed systems, human-computer interaction, and model-driven engineering. Each of these domains presents both a research opportunity and a pathway to tangible healthcare impact. By tackling these issues directly, software engineers can help PMDTs mature from visionary prototypes into the foundation of proactive, personalized, and trustworthy chronic care.

# LESSONS LEARNED FROM EARLY PILOTS

While the Patient Medical Digital Twin (PMDT) remains a visionary paradigm, early piloting efforts have already generated valuable insights into its feasibility, clinical relevance, and engineering requirements. Our work has progressed along three complementary strands: ontology-driven modeling[13], federated analytics and privacy-preserving AI[14], and model-driven engineering with domain-specific languages[15]. Collectively, these experiences highlight both the potential of PMDTs and the key challenges that must be addressed for large-scale, sustainable adoption.

**Ontology-Driven Integration is Feasible but Iterative:** The QUALITOP oncology pilot demonstrated that modular blueprints can effectively represent multimodal patient data, including clinical, psychosocial, and lifestyle information, within a unified ontology. Validation with clinical experts confirmed both feasibility and clinical relevance. However, alignment with standards such as HL7 FHIR and OMOP proved to be an iterative process rather than a one-time mapping exercise. Achieving semantic consistency required a repeated refinement and pragmatic compromises, including the use of bridging ontologies and partial alignments, to ensure usability across heterogeneous hospital systems. These insights are elaborated in[13]

**Federated Analytics Strengthen Trust but Add Engineering Overhead:** The Smart Digital Health Platform presented in[14], operationalized PMDT concepts within a distributed environment. Federated queries across hospital datasets successfully produced predictive insights without centralizing sensitive data, thereby reinforcing GDPR and HIPAA compliance. However, these pilots also revealed significant engineering overhead associated with managing provenance metadata, maintaining consistent semantics across institutions, and preserving performance under load. These findings underline that governance-aware federated infrastructures are not optional but essential. Their design must explicitly address latency, versioning, and institutional heterogeneity to ensure reliability and trust at scale.

**Usability Is as Critical as Semantics:** Across both ontology-driven and federated pilots, workshops with clinicians confirmed that PMDTs could answer competency questions (e.g., predicting risk factors for adverse events), thus strengthening confidence in the underlying models. Yet, interfaces often proved too complex for routine clinical use. This finding emphasized the need for co-design of visualizations, dashboards, and explanatory tools with clinicians and patients to balance transparency and cognitive load. Usability emerged as a determinant of adoption equal in importance to interoperability and privacy, reinforcing that technical sophistication alone is insufficient for clinical impact.

**DSL-Driven Processing Enhances Accessibility:** To address system complexity, we are developing the Digital

Twin Processing Language (DTPL), a domain-specific language within a model-driven engineering framework[15]. Early prototypes showed how high-level abstractions can make complex analytics more accessible, enabling descriptive and predictive "what-if" queries such as simulating treatment alternatives. Stakeholder feedback emphasized strong demand for such capabilities, confirming the importance of DSL/MDE approaches for both scalability and accessibility. Preliminary results suggest that DTPL can elevate PMDTs from static repositories to configurable, adaptive entities capable of evolving alongside clinical knowledge and practice.

**Adoption Depends on Demonstrable Value:** Perhaps the most important lesson across all pilots is that adoption depends not on technical novelty but on measurable improvements in clinical workflows, patient outcomes, and policy decision-making. PMDTs must prove that they deliver tangible value: streamlining workflows, improving patient quality of life, or enabling more informed data-driven interventions. Embedding evaluation metrics for usability, trust, and clinical impact from the outset is therefore essential for real-world translation and sustained adoption.

**Summary.** Early pilots confirm that PMDTs are both technically feasible and clinically meaningful, but they also expose persistent gaps in interoperability, governance, usability, and abstraction. These lessons reinforce that engineering Patient Medical Digital Twins is a socio-technical endeavor, requiring iterative alignment with evolving standards, trust-preserving governance mechanisms, and human-centered design. At the same time, they outline clear directions for progress: scalable interoperability frameworks, governance-aware federated infrastructures, and model-driven abstractions such as DTPL.

The next section builds upon these insights to articulate a research and practice agenda for software engineers to advance PMDTs from promising pilot implementations toward sustainable, ecosystem-wide adoption.

## THE ROAD AHEAD

Although Patient Medical Digital Twins (PMDTs) remain in their infancy, early pilots confirm both their feasibility and the challenges that must be overcome for sustainable adoption. Moving from research prototypes to operational healthcare systems will require deliberate advances in software engineering. Based on our experience in ontology-driven modeling, federated analytics, and emerging DSL/MDE approaches, we identify five key directions for future progress.

**From Prototypes to Platforms:** Current PMDT implementations are fragmented, typically confined to single pilots or disease-specific contexts. Achieving broader impact will require evolving toward robust, reusable platforms that can support multiple chronic conditions and adapt to evolving interoperability standards. Software engineers must design architectures emphasizing modularity, versioning, and maintainability at platform scale, ensuring that PMDTs transition from isolated experiments to deployable, interoperable assets across healthcare systems.

**Governance by Design:** Trust will be the decisive factor undermining PMDT adoption. Governance cannot remain an afterthought; it must be embedded as a core architectural principle. This entails formalizing provenance, consent, and access control policies and machine-actionable models and engineering infrastructures that reason automatically about permissible data sharing under frameworks such as GDPR and HIPAA. Future research should explore privacy-enhancing technologies, including secure multiparty computation, homomorphic encryption, and blockchain, while also developing transparent, auditable, and clinically interpretable policy models that preserve both compliance and user trust.

**Human-Centered PMDTs:** Pilot evaluations consistently demonstrate that usability is as critical as semantic rigor. The next generation of PMDT interfaces must be co-designed with clinicians and patients to deliver explainable AI, meaningful visualizations, and interactive dashboards that translate raw data into actionable insights. For patients, this implies digital companions that facilitate self-management without technical complexity; for clinicians, decision-support tools that integrate seamlessly into established workflows. Research into interaction design, cognitive ergonomics, and explainable AI for PMDTs will therefore be as decisive as advances in data integration and analytics.

**Programmability Through DSLs and MDE:** Managing PMDT complexity requires abstractions that make systems configurable, analyzable, and evolvable by design. Our ongoing work on the Digital Twin Processing Language (DTPL) illustrates this potential: descriptive and predictive queries, blueprint composition for multimorbidity, and scenario-based simulations can be expressed at a high level and automatically transformed into executable components. Future work should pursue co-design of DSLs with clinical and data domain experts to ensure that languages remain both expressive and intuitive, and validate them through real-world clinical pilots to confirm usability, correctness, and domain alignment.

**Toward Learning Health Ecosystems:** The most ambitious opportunity lies in scaling from individual twins to ecosystem-wide infrastructures that learn collectively. Aggregated PMDTs can drive preventive health strategies, optimize resource allocation, and generate population-level insights. Realizing this vision will require advances in federated learning, ontology-driven reasoning, and governance-aware orchestration, supported by embedded evaluation metrics for clinical impact, usability, and societal trust. In such learning ecosystems, PMDTs will serve not only as digital representation of patients bus as active computational nodes in continuously improving, data-driven health systems.

**Summary:** The path forward for PMDTs is not solely a clinical or data science challenge, it is fundamentally a software engineering challenge. Addressing interoperability, governance, usability, model-driven programmability, and distributed learning will be essential to transform PMDTs from early research pilots into the foundation of proactive, personalized, and trustworthy chronic care. The opportunity is unprecedented: to transition healthcare from reactive treatment toward adaptive, continuously optimized, and universally accessible health management.

# CONCLUSION

Patient Medical Digital Twins (PMDTs) represent more than an aspirational concept; they mark a paradigm shift in how chronic care can be modeled, governed, and delivered. Early pilot studies demonstrate that ontology-driven modeling is feasible, that federated analytics can preserve privacy while enabling predictive insights, and that domain-specific abstractions like DSLs can make complexity tractable. These results confirm that PMDTs are not speculative constructs but emerging realities within clinical settings. Their continued success, however, depends on strengthening the software engineering foundations that support them

The challenges ahead are clear: achieving interoperability across heterogeneous standards; embedding governance mechanisms that formalize provenance, consent, and policy reasoning; designing scalable architectures that evolve alongside care pathways; and pursuing human-centered design that transforms digital twins into trusted companions rather than technical curiosities. Meeting these challenges requires the same rigor, abstraction, and systematic thinking that define the discipline of software engineering.

The opportunity, however, is unprecedented. If realized, PMDTs could transform healthcare from reactive treatment to proactive, adaptive, and sustainable care, empowering patients through lifelong digital companions, supporting clinicians with actionable intelligence, and enabling health systems to learn continuously at scale. For the software engineering community, PMDTs represent not only a new frontier in healthcare innovation but also a defining opportunity to demonstrate how trustworthy, adaptive, and human-centered digital infrastructures can be engineered for the benefit of society as a whole.

## ACKNOWLEDGMENTS
We thank Hospital Clinic Barcelona (IDIBAPS), Hospices Civils de Lyon (HCL), University Medical Center Groningen (UMCG), and Instituto Português de Oncologia, Lisboa (IPOL) for their contributions to the pilot study and their continuous support in evaluation and validation of the reported results. We also thank Dr. Roxana Albu, Chief Scientific Officer at the Association of European Cancer Leagues (ECL), and Dr. Menia Koukougianni, Fellow of the European Patients' Academy and Baheya Hospital in Egypt, for their valuable feedback in assessing this work. This work was supported in part by the European Commission Horizon 2020 project QUALITOP under Grant H2020-SC1-DTH-01-2019-875171.

**Michael P. Papazoglou** is the co-founder and Vice President of the Scientific Academy for Service Technology (ServTech). His current research interests include: Emerging technologies, Industrial engineering, Smart Applications and Smart Technology Solutions for Healthcare and Manufacturing. Papazoglou received the Ph.D. in Computer Systems Engineering from the University of Dundee. He is a Fellow of the IEEE Computer Society. Contact him at mikep@servtech.info.

**Bernd J. Krämer** is the co-founder and president of the Scientific Academy for Service Technology (ServTech). He is a Professor Emeritus of Software Engineering at FernUniversität in Hagen, Germany. His current research interests include: Emerging technologies, Industrial engineering, Smart Applications and Smart Technology Solutions for Healthcare and Manufacturing. Krämer received the Ph.D. in electrical engineering and computer science from the Technical University of Berlin. He is a Fellow of the IEEE Computer Society. Contact him at kraemer@servtech.info.

**Mira Raheem** is a teaching assistant at the Faculty of Computers and Artificial Intelligence, Cairo University, Egypt, and a visiting researcher at the Scientific Academy for Service Technology (ServTech), Potsdam, Germany. Her research interests include model-driven engineering, artificial intelligence in healthcare, and semantic interoperability. She holds a B.Sc. in information systems from Cairo University. Contact her at mira.raheem@fci.cu.edu.eg.

**Amal Elgammal** is head of the Software Engineering Department at Egypt University of Informatics and an associate professor of software engineering at Cairo University, Egypt. She is also a senior visiting researcher at the Scientific Academy for Service Technology (ServTech). Her research interests include emerging technologies, industrial engineering, smart applications, and technology solutions for healthcare and manufacturing. Elgammal received a Ph.D. in information management from Tilburg University, the Netherlands. Contact her at amal@servtech.info.